\documentclass[prd,twocolumn, nofootinbib]{revtex4}

\pdfoutput=1

\usepackage{graphicx}
\usepackage{subfig}

\usepackage{float}
\usepackage{enumerate}
\usepackage{amssymb,amsmath,bm}  

\usepackage[usenames,dvipsnames]{xcolor}
\usepackage{hyperref}
\hypersetup{
    colorlinks = true,
    citecolor = {MidnightBlue},
    linkcolor = {BrickRed},
    urlcolor = {BrickRed}
}
\usepackage{lipsum}  
\usepackage{hyperref}

\newcommand{\ba}{\begin{eqnarray}}
\newcommand{\ea}{\end{eqnarray}}
\begin{document}
\title{Cosmology with Planck T-E correlation coefficient}
\date{}
\author{A. La Posta}\affiliation{Université Paris-Saclay, CNRS, IJCLab, 91405, Orsay, France}
\author{T. Louis}\affiliation{Université Paris-Saclay, CNRS, IJCLab, 91405, Orsay, France}
\author{X. Garrido}\affiliation{Université Paris-Saclay, CNRS, IJCLab, 91405, Orsay, France}
\author{M. Tristram}\affiliation{Université Paris-Saclay, CNRS, IJCLab, 91405, Orsay, France}
\author{S. Henrot-Versill\'e}\affiliation{Université Paris-Saclay, CNRS, IJCLab, 91405, Orsay, France}

\begin{abstract}

Tensions in cosmological parameters measurement motivate a revisit of the effects of instrumental systematics. In this article, we focus on the Pearson’s correlation coefficient of the cosmic microwave background temperature and polarization E modes $\mathcal{R}_\ell^{\rm TE}$ which has the property of not being biased by multiplicative instrumental systematics. We build a $\mathcal{R}_\ell^{\rm TE}$-based likelihood for the Planck data, and present the first constraints on $\Lambda$CDM parameters from the correlation coefficient. Our results are compatible with parameters derived from a power spectra based likelihood. In particular the value of the Hubble parameter $H_0$ characterizing the expansion of the Universe today, 67.5 $\pm$ 1.3 km/s/Mpc, is consistent with the ones inferred from standard CMB analysis. We also discuss the consistency of the Planck correlation coefficient with the one computed from the most recent ACTPol power spectra.

\end{abstract}

  \date{\today}
  \maketitle

\section{Introduction}\label{sec:Intro}

Increasingly precise constraints on $\Lambda$CDM parameters from different probes have revealed tensions between measurements from early and late time Universe, including the well-known $H_0$ tension. The value of $H_0$ inferred from Cosmic Microwave Background (CMB) temperature, polarization and lensing anistropies by the Planck Collaboration is 67.36 $\pm$ 0.54 km/s/Mpc~\cite{Planck2018:cosmo}. High resolution CMB observations from ground based experiments such as ACTPol and SPT3G have also independently measured $H_0$ and found 67.9 $\pm$ 1.5 km/s/Mpc~\cite{ACT:A20} and 68.8 $\pm$ 1.5 km/s/Mpc~\cite{dutcher2021measurements} respectively.  All of these measurements are inconsistent with the latest measurement from the Cepheids-calibrated cosmic distance ladder $H_0 = 73.2 \pm 1.3$ km/s/Mpc~\cite{Riess2021}.\\

This discrepancy could indicate the need for new physics beyond $\Lambda$CDM. Many models have already been proposed to solve the Hubble tension. Modifications to early time physics, such as new physics that change the physical size of the sound horizon at recombination, are investigated and the different ways of solving the current tension have been reviewed in~\cite{Knox2020, Divalentino2021}. Upcoming cosmic variance limited measurements of CMB polarization for a wide multipole range from Simons Observatory~\cite{Ade2019} and CMB-S4~\cite{Abazajian2016} will put severe constraints on these models.\\

Another hypothesis which could explain these tensions is the presence of remaining systematic effects not accounted for in the data analysis. The presence of bias in the cosmic distance ladder data analysis is actively studied, in particular biases coming from supernovae light curves standardization~\cite{Rigault2018, Saunders2018}. An alternative calibration of cosmic distances, using the Tip of the Red Giant Branch (TRGB) is also explored, giving another measurement of the Hubble parameter~\cite{Freedman2019, Freedman2020}.\\

In this paper we focus on possible instrumental systematic effects that can affect the measurement of CMB anisotropies. A way of testing this possibility is to use observables that are less sensitive to the exact instrument model. One of these observables is the Pearson's correlation coefficient of T and E modes $\mathcal{R}_\ell^{\rm TE}=C_\ell^\mathrm{TE} / \sqrt{C_\ell^\mathrm{TT}C_\ell^\mathrm{EE}}$. It was mentioned in~\cite{Planck2015:ps}, in which the authors give geometrical interpretations of the correlation coefficient as a cosine or as a decorrelation angle. Correlation coefficients have also been used to look at the correlations between galaxy surveys and CMB-derived lensing power spectrum in an unbiased way~\cite{Peacock2018}.  $\mathcal{R}_\ell^{\rm TE}$ statistical properties have been discussed in details in~\cite{Louis2019}. This observable has been shown to be robust against multiplicative instrumental systematics, such as beam error, calibration, polarization efficiency uncertainties or unmodelled transfer functions.\\

This work presents the first estimation of the cosmological parameters from $\mathcal{R}_\ell^{\rm TE}$. The nearly Gaussian nature of the correlation coefficient, at scales where we can ensure a high $EE$ signal-to-noise ratio, makes possible the use of a Gaussian likelihood function to constrain cosmological parameters and to compare with constraints from the combination of $C_\ell^{\rm TT}$, $C_\ell^{\rm TE}$ and $C_\ell^{\rm EE}$ power spectra.\\

The paper is organized as follows. In section~\ref{sec:syst} we illustrate the effect of different systematic effects on cosmological parameters constrained from the CMB power spectra. In section~\ref{sec:like} we discuss the construction of a $\mathcal{R}_\ell^{\rm TE}$-based likelihood. In section~\ref{sec:data} we show the results of a $\mathcal{R}_\ell^{\rm TE}$-based analysis for the Planck PR4 dataset. We conclude in section~\ref{sec:Conclusion}.

\section{Effect of systematics in power spectra based likelihood}\label{sec:syst}

In this section, we introduce the dataset and likelihood used in the rest of the paper, we then study the impact of systematics on cosmological parameter estimation using a set of biased simulations.

\subsection{Dataset and $C_\ell$-based likelihood}\label{subsec:dataset}

We use the latest Planck data release (PR4). The maps were produced using the \texttt{NPIPE} processing pipeline that jointly analyzes data from Planck High Frequency Instrument (HFI) and Planck Low Frequency Instrument (LFI)~\cite{PlanckNPIPE}.\\

Cosmological parameters are estimated using the Planck \texttt{HiLLiPoP} (High-L Likelihood Polarized for Planck) likelihood. The code is publicly available on github\footnote{\url{https://github.com/planck-npipe}}. This is a multifrequency likelihood for the Planck cosmology channels : 100, 143 and 217 GHz. The dataset consists of two split map for each frequency. From these maps, we obtain 15 cross power spectra from which we compute the six cross-frequency spectra used in the likelihood. The foreground model includes galactic dust, CIB, tSZ \& kSZ contributions, tSZ-CIB correlation and Poisson-like distributed point source foregrounds. A more detailled description is given in~\cite{Couchot2017}. The likelihood assumes that the $TT$, $TE$ and $EE$ CMB power spectra follow a Gaussian distribution, which is a good assumption at high multipoles ($\ell>30$). The likelihood has been tested extensively against the Plik likelihood~\cite{Planck2015:ps}.\\

In section~\ref{subsec:syst}, we study the impact of systematics on simulated Planck data. Cosmological constraints from the Planck PR4 dataset are presented in section~\ref{sec:data}.
\subsection{Bias on the cosmological parameters}\label{subsec:syst}
The correlation coefficient is designed to be insensitive to any multiplicative bias. To illustrate the effect of this kind of bias on cosmological parameters estimated from the combination of the $TT$, $TE$ and $EE$ power spectra, we generate a set of biased simulations and fit for the cosmological parameters. The simulations are generated at the power spectra level from the best fit \texttt{HiLLiPoP} cosmology and foreground model, we then use the Planck PR4 covariance matrix to add scatter to the spectra. Fiducial cosmology has been set to $100\theta_{\mathrm{MC}} = 1.04065$, $\Omega_bh^2 = 0.02231$, $\Omega_ch^2 = 0.1193$, $\mathrm{ln}(10^{10}A_\mathrm{s}) = 3.045$, $n_\mathrm{s} = 0.9619$ and $\tau = 0.0566$.\\

We include systematics into our simulated dataset such that the observed temperature $\Tilde{a}_{\ell m}^\mathrm{T}$ and polarization E-modes $\Tilde{a}_{\ell m}^\mathrm{E}$ are given by $\Tilde{a}_{\ell m}^\mathrm{T} = \epsilon_\ell^\mathrm{T}a_{\ell m}^\mathrm{T}$ and $\Tilde{a}_{\ell m}^\mathrm{E} = \epsilon_\ell^\mathrm{P}a_{\ell m}^\mathrm{E}$. The measured power spectra are
\begin{align}\label{eq:bias}
    \Tilde{C}_\ell^{TT} &= ({\epsilon_\ell^{\mathrm{T}}})^2C_\ell^{TT}\nonumber\\
    \Tilde{C}_\ell^{TE} &= {\epsilon_\ell^{\mathrm{T}}}\epsilon_\ell^{\mathrm{P}}C_\ell^{TE}\nonumber\\
    \Tilde{C}_\ell^{EE} &= ({\epsilon_\ell^{\mathrm{P}}})^2C_\ell^{EE}
\end{align}
Systematic effects can be constant over multipoles (e.g. polarization efficiency) or scale-dependant (e.g. transfer functions).\\

\begin{figure*}
\centering
\subfloat[Polarization efficiency]{\label{fig1:poleff}\includegraphics{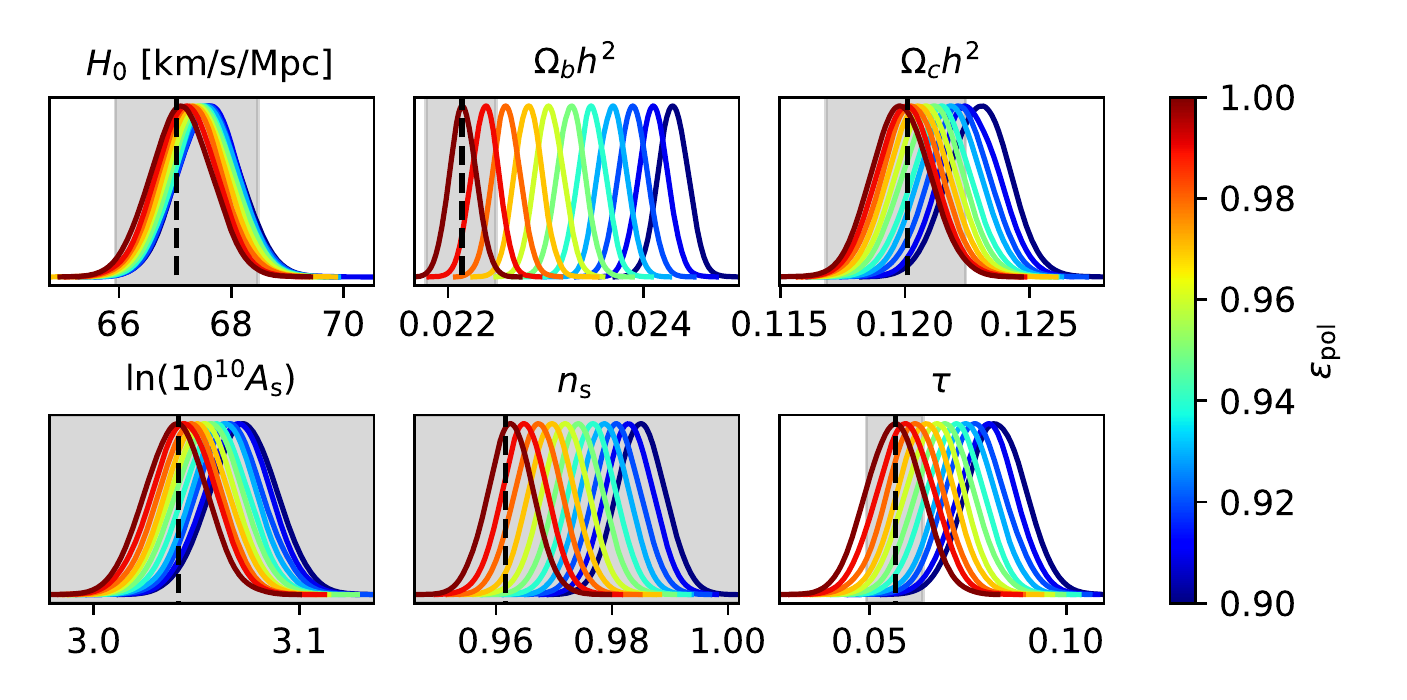}}

\subfloat[Polarization transfer function]{\label{fig1:polTF}\includegraphics{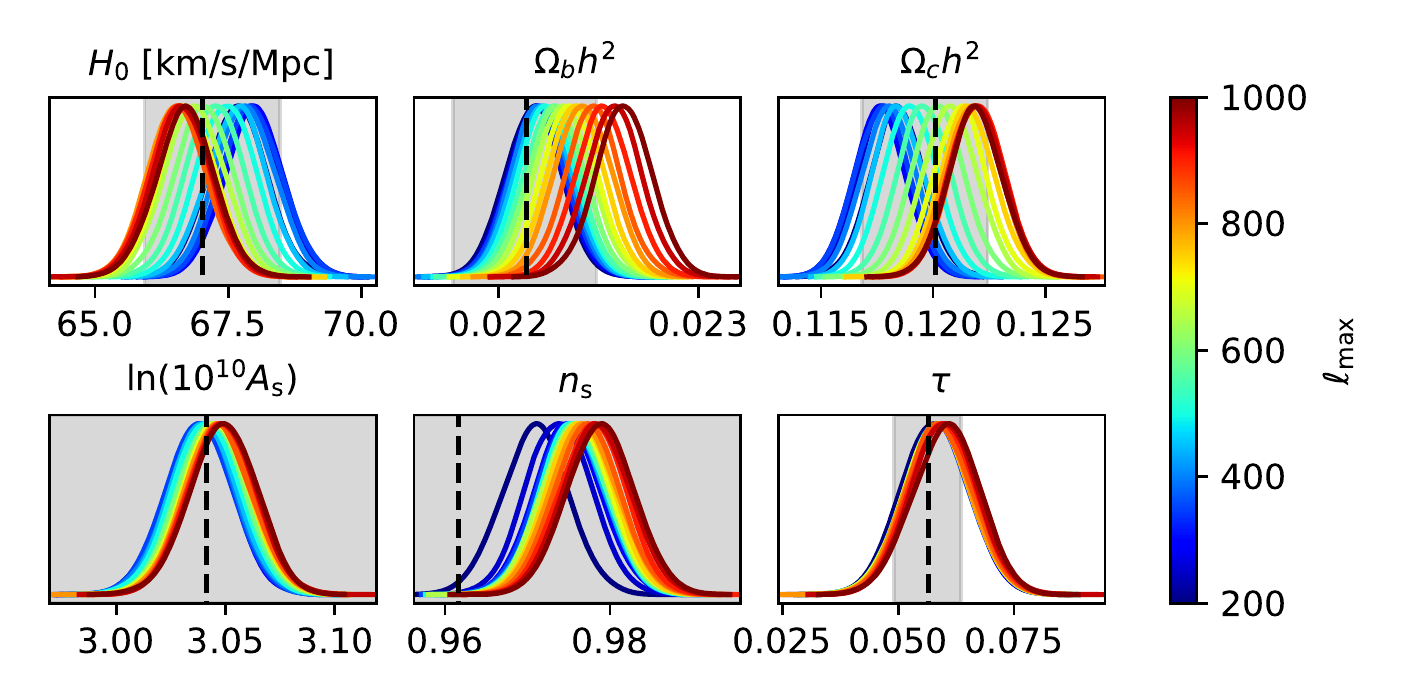}}

\subfloat[Temperature transfer function]{\label{fig1:tempTF}\includegraphics{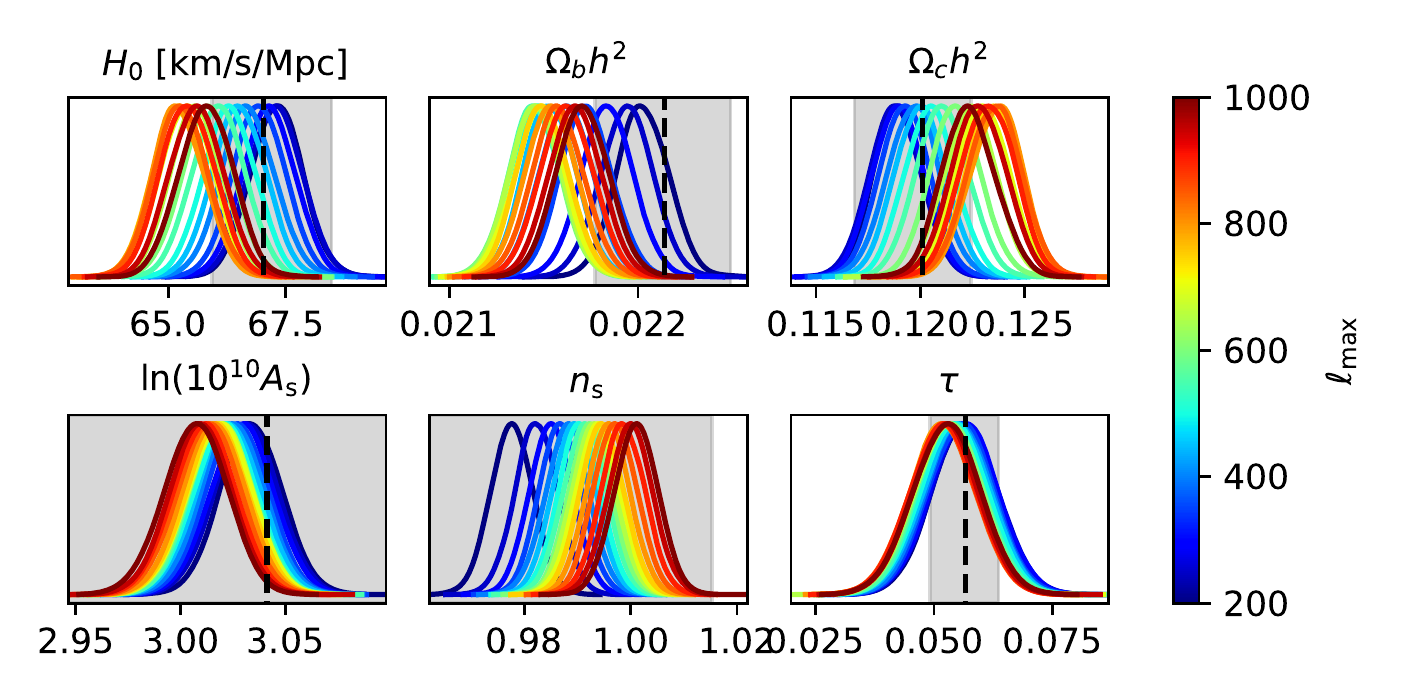}}
\caption{\label{fig:bias_syst} 1D posterior distributions of the six $\Lambda$CDM parameters using a simulated dataset ($TT$, $TE$, $EE$) biased with a constant polarization efficiency~(\ref{fig1:poleff}), with a polarization transfer function~(\ref{fig1:polTF}) and with a temperature transfer function~(\ref{fig1:tempTF}). The transfer function model is described in Eq.~(\ref{eq:tf_model}). We vary the~$\ell_{\rm max}$ parameter from $200$ to $1000$. The dashed black lines correspond to the fiducial model. Gray bands correspond to $\pm 1\sigma$ $\mathcal{R}_\ell^{\rm TE}$ constraints.}
\end{figure*}

We obtain the posterior distributions of cosmological parameters using the Markov Chain Monte Carlo (MCMC) algorithm implemented in \texttt{cobaya}~\cite{Torrado2020} with the $C_\ell$-based \texttt{HiLLiPoP} likelihood. Figure~\ref{fig:bias_syst} displays the distributions of the six standard $\Lambda$CDM parameters for a set of different transfer functions. We consider three different kinds of systematics : 
\begin{enumerate}[a.]
    \item $\epsilon^{\rm P}_\ell = \mathrm{cte}$, $\epsilon^{\rm T}_\ell = 1$ (Fig.~\ref{fig1:poleff})
    \item $\epsilon^{\rm P}_\ell = \epsilon^{\rm P}(\ell)$, $\epsilon^{\rm T}_\ell = 1$ (Fig.~\ref{fig1:polTF})
    \item $\epsilon^{\rm P}_\ell = 1$, $\epsilon^{\rm T}_\ell = \epsilon^{\rm T}(\ell)$ (Fig.~\ref{fig1:tempTF})
\end{enumerate}

For illustration purpose we choose a transfer function such as
\begin{equation}\label{eq:tf_model}
    \epsilon(\ell) = 
    \begin{cases}
        \epsilon_{\mathrm{min}} & \text{if $\ell < \ell_{\mathrm{min}}$} \\
        \epsilon_{\mathrm{min}} + \Delta\epsilon\cdot\mathrm{sin}^2\left(\frac{\pi}{2}\frac{\ell - \ell_{\mathrm{min}}}{\Delta\ell}\right) & \text{if $\ell_{\mathrm{min}}\le\ell\le\ell_{\mathrm{max}}$} \\
        \epsilon_{\mathrm{max}} & \text{if $\ell > \ell_{\mathrm{max}}$}
    \end{cases}
\end{equation}
where $\Delta\epsilon = \epsilon_{\rm max} - \epsilon_{\rm min}$ and $\Delta\ell = \ell_{\rm max} - \ell_{\rm min}$ with $\ell_{\rm min} = 100$. The transfer function will smoothly increase from $\epsilon_\mathrm{min} = 0.95$ to $\epsilon_\mathrm{max} = 1$ between $\ell_\mathrm{min}$ and $\ell_\mathrm{max}$.\\

Adding an inconsistency between temperature and polarization leads to significant shifts for all the cosmological parameters for the three different models used in this analysis. The effect is clearly noticeable for baryon density and particularly for the polarization efficiency case~(\ref{fig1:poleff}). Using the biased simulations, we obtain constraints on $\Omega_bh^2$ that are more than $3\sigma$ discrepant with the unbiased $C_\ell$s-derived constraint for $\epsilon_{\rm P}\le0.97$. $H_0$ is particularly affected by a temperature transfer function : the $H_0$ constraint is shifted towards lower values~(\ref{fig1:tempTF}). The $\mathcal{R}_\ell^{TE}$ correlation coefficient could therefore be an interesting consistency test, it could help avoiding large bias in our cosmological parameter measurements arising from multiplicative systematic effects.\\

\section{A likelihood for the Pearson's correlation coefficient of T and E modes}\label{sec:like}

In this section we recall some of the statistical properties of $\mathcal{R}_\ell^{\rm TE}$ and we propose and validate a multifrequency likelihood for estimating cosmological parameters based on the correlation coefficients.
\begin{figure*}
\includegraphics[scale = 0.9]{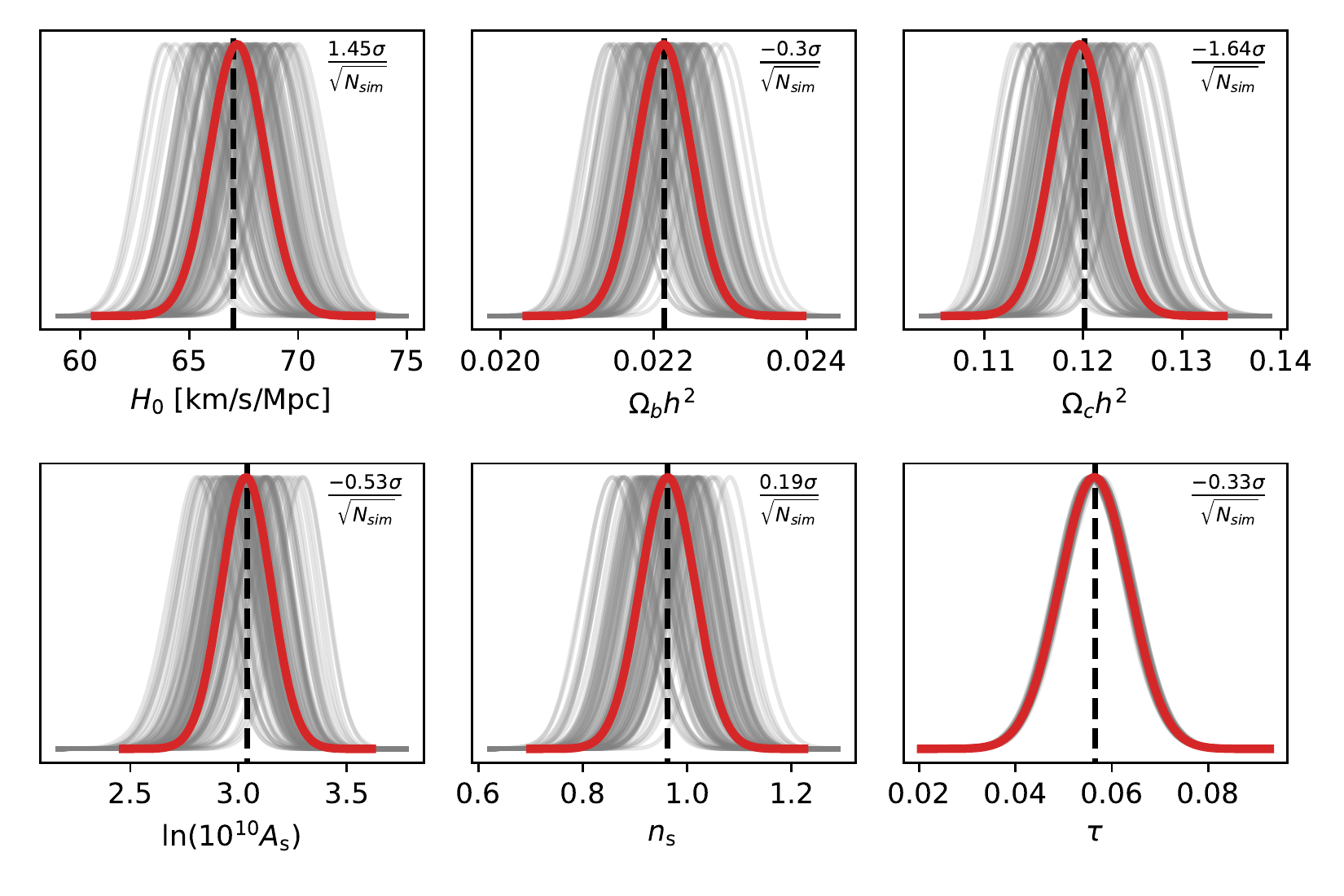}
\caption{\label{fig:check_like} $\Lambda$CDM parameters constraints for a set of $N_{\rm sim}=100$ simulations (gray). The average posterior distribution is shown in red. The distance from the average to the input value (in sigma units) is displayed in the top right corner of each panel. Mean posterior distributions are consistent with input values of the simulations}
\end{figure*}
\subsection{Statistic of the multifrequency correlation coefficient}\label{subsec:stat}
An estimator for the correlation coefficient can be obtained using the estimated $TT$, $EE$ and $TE$ power spectra : $\hat{\mathcal{R}}_b^{TE} = {\hat{C}_b^{TE}}/{\sqrt{\hat{C}_b^{TT}\hat{C}_b^{EE}}}$. By definition, the measured $\Tilde{\mathcal{R}}_\ell^{\rm TE}$ correlation coefficient is unaffected by any of the systematic effects we discuss in section~\ref{sec:syst}. (i.e $\Tilde{\mathcal{R}}_\ell^{\rm TE} = \mathcal{R}_\ell^{\rm TE}$).\\

This estimator is affected by a subdominant bias that can be corrected at first order as : $\hat{\mathcal{R}}_b^{\mathrm{TE},c} = \hat{\mathcal{R}}_b^{\mathrm{TE}}(1-\alpha_b)$. An analytical expression for $\alpha_b$ is given in appendix~\ref{app:statistics}. The expression for the covariance matrix of the $\mathcal{R}_b^{\rm TE}$ estimator was also derived in~\cite{Louis2019}, the generalization of the expression to a multifrequency case is given by
\begin{align}\label{eq:covmat}
    \mathbf{\Gamma}(\mathcal{R}_b^{\rm TE}, \mathcal{R}_{b'}^{\rm TE}) &= \mathbf{\Gamma}(C_{b}^{\rm{TE}}, C_{b'}^{\rm{TE}})\nonumber\\
    &+ \frac{1}{4}\big[\mathbf{\Gamma}(C_{b}^{\rm{TT}}, C_{b'}^{\rm{TT}}) + \mathbf{\Gamma}(C_{b}^{\rm{EE}}, C_{b'}^{\rm{EE}})\big]\nonumber\\
    &- \frac{1}{2}\big[\mathbf{\Gamma}(C_{b}^{\rm{TE}}, C_{b'}^{\rm{TT}}) + \mathbf{\Gamma}(C_{b}^{\rm{TT}}, C_{b'}^{\rm{TE}})\nonumber\\
    &\quad+ \mathbf{\Gamma}(C_{b}^{\rm{TE}}, C_{b'}^{\rm{EE}}) + \mathbf{\Gamma}(C_{b}^{\rm{EE}}, C_{b'}^{\rm{TE}})\big]\nonumber\\
    &+ \frac{1}{4}\big[\mathbf{\Gamma}(C_{b}^{\rm{TT}}, C_{b'}^{\rm{EE}}) + \mathbf{\Gamma}(C_{b}^{\rm{EE}}, C_{b'}^{\rm{TT}})\big]
\end{align}
where $\mathbf{\Gamma}(\rm{X}, \rm{Y}) = {\text{cov}(X^{\nu_1\times\nu_2}, Y^{\nu_3\times\nu_4})}/{(X^{\nu_1\times\nu_2}\cdot Y^{\nu_3\times\nu_4})}$. Derivation of Eq.~(\ref{eq:covmat}) is developed in appendix~\ref{app:statistics}.

\subsection{$\mathcal{R}_\ell^{\rm TE}$-based likelihood construction}\label{subsec:Rlike}
We adapt the \texttt{HiLLiPoP} $C_\ell$-based likelihood introduced in section~\ref{sec:syst} in order to fit cosmological parameters from the correlations coefficient. We use the multi-frequency covariance matrix given in Eq.~(\ref{eq:covmat}) and assume that the correlation coefficients follow a multivariate Gaussian distribution.\\

\noindent The likelihood is defined as follows : 
\begin{align}\label{eq:like}
    \mathrm{ln}\mathcal{L} \simeq &-\frac{1}{2}\left(\Delta\mathcal{R}^{\mathrm{vec}}\right)^{\mathrm{T}}\mathbf{\Xi}^{-1}\left(\Delta\mathcal{R}^{\mathrm{vec}}\right)
\end{align}
where $\mathbf{\Xi}$ is the $\mathcal{R}_\ell^{TE}$ multi-frequency covariance matrix and $\Delta\mathcal{R}^{\mathrm{vec}} = \mathcal{R}^{\mathrm{vec},\mathrm{data}} - \mathcal{R}^{\mathrm{vec},\mathrm{th}}$ is the residual correlation coefficient vector including the 6 cross-frequency spectra. The $\mathcal{R}_\ell^{\rm TE}$ covariance matrix is computed from the Planck PR4 power spectra covariance matrices using Eq.~(\ref{eq:covmat}).\\
\begin{figure*}
\centering
\includegraphics{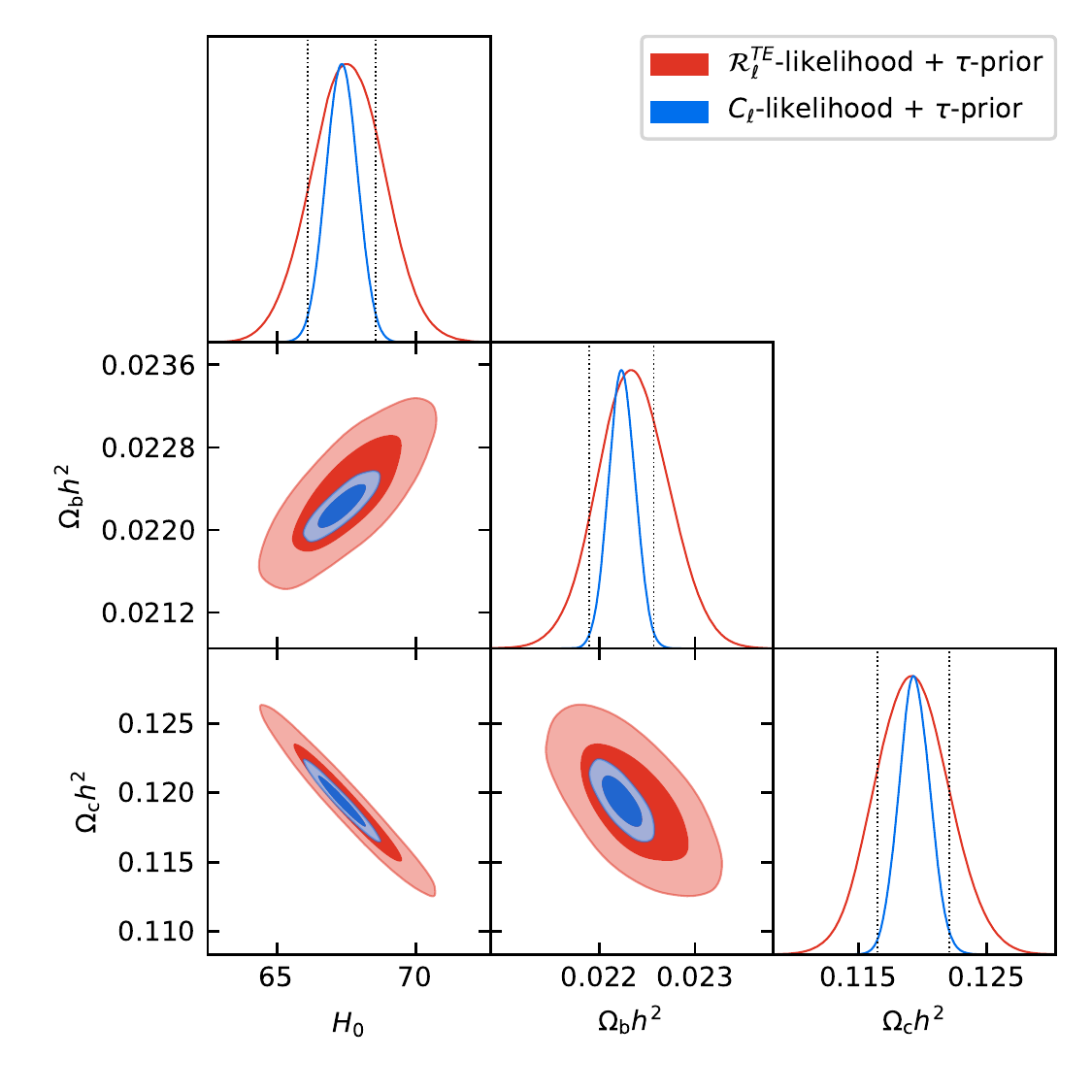}
\caption{\label{fig:cosmo} 2D posterior distributions for $H_0$, $\Omega_bh^2$ and $\Omega_ch^2$ constrained using $C_\ell$-based likelihood (blue) or the $\mathcal{R}_\ell^{TE}$-based likelihood (red). The expected 1$\sigma$ scatter between parameters estimated from the two different likelihoods is displayed with gray dotted lines.}
\end{figure*}

The data vector is computed from the $C_\ell$s data vector using the unbiased estimator $\hat{\mathcal{R}}_b^{\mathrm{TE},c}$. The model for the correlation coefficient is constructed using theoretical and foregrounds $TT$, $EE$, and $TE$ power spectra. For a cross frequency $\nu_1\times\nu_2$, we model the $XY$ power spectra as
\begin{align}\label{eq:model_like}
    C_\ell^{XY, \mathrm{th},\nu_1\times\nu_2} = A_{\rm Pl}A^{XY, \nu_1\times\nu_2}[&C_\ell^{XY, \mathrm{CMB}}(\theta)\nonumber\\ &+ C_\ell^{XY, \mathrm{fg}, \nu_1\times\nu_2}(\theta_{\rm fg})]
\end{align}
where $(X, Y)\in\{T, E\}^2$, $A^{XY, \nu_1\times\nu_2}$ depends on the map calibration parameters and $A_{\rm Pl}$ is a global calibration parameter. $\theta_{\rm fg}$ is the set of parameters describing the foregrounds power spectra amplitudes and $\theta$ are the $\Lambda$CDM parameters. We give more details about Eq.~(\ref{eq:model_like}) in appendix~\ref{app:ps_model}.\\

\begin{table*}[t]
\begin{ruledtabular}
\begin{tabular}{lp{1.5cm}p{1.5cm}p{1.5cm}p{1.5cm}p{1.5cm}p{1.5cm}||p{1.5cm}p{1.5cm}}
 &\multicolumn{2}{c}{$C_\ell^{\rm TT}$}&\multicolumn{2}{c}{$C_\ell^{\rm EE}$}&\multicolumn{2}{c||}{$C_\ell^{\rm TE}$}&\multicolumn{2}{c}{$\mathcal{R}_\ell^{\rm TE}$}\\
Frequencies (GHz) & \hfil$\ell_{\mathrm{min}}$ & \hfil$\ell_{\mathrm{max}}$ & \hfil$\ell_{\mathrm{min}}$ & \hfil$\ell_{\mathrm{max}}$ & \hfil$\ell_{\mathrm{min}}$ & \hfil$\ell_{\mathrm{max}}$ & \hfil$\ell_{\mathrm{min}}$ & \hfil$\ell_{\mathrm{max}}$\\
\hline
    100$\times$100 &  \hfil30 & \hfil1200 & \hfil30 & \hfil1000 & \hfil30 & \hfil1200 & \hfil50 & \hfil1000 \\ 
    100$\times$143 & \hfil30 & \hfil1500 & \hfil30 & \hfil1250 & \hfil30 & \hfil1500 & \hfil50 & \hfil1250 \\
    100$\times$217 & \hfil100 & \hfil1500 & \hfil400 & \hfil1250 & \hfil300 & \hfil1500 & \hfil400 & \hfil1250 \\
    143$\times$143 & \hfil30 & \hfil2000 & \hfil30 & \hfil1750 & \hfil30 & \hfil1750 & \hfil50 & \hfil1500 \\
    143$\times$217 & \hfil100 & \hfil2500 & \hfil400 & \hfil1750 & \hfil300 & \hfil1750 & \hfil400 & \hfil1500 \\
    217$\times$217 & \hfil100 & \hfil2500 & \hfil400 & \hfil2000 & \hfil300 & \hfil2000 & \hfil400 & \hfil1500\\
\end{tabular}
\end{ruledtabular}
\caption{\label{tab:multipoles}Multipole ranges used for each power spectrum and for the $\mathcal{R}_\ell^{\rm{TE}}$ correlation coefficient.}
\end{table*}

While the approximation of Gaussianity is excellent for CMB power spectra at multipoles $\ell \ge 30$, it is not straightforward that it holds for a non linear combination of power spectra. However, the Gaussianity of the correlation coefficient has been shown to be a good assumption in the high $EE$ signal-to-noise regime~\cite{Louis2019}. We check the robustness of this approximation running MCMC chains for the set of simulations described in section~\ref{sec:syst}. Figure~\ref{fig:check_like} displays the distributions and the mean posterior for each $\Lambda$CDM parameters for $N_{\rm sim} = 100$ simulations. Recovered posterior distributions are statistically consistent with expectations. The use of a $\tau$-prior explains the small dispersion of the posterior distributions for this parameter.

\section{Application to Planck data}\label{sec:data}

In this section, we estimate cosmological parameters using the $\mathcal{R}_\ell^{\rm TE}$-likelihood on Planck PR4 data. We also compute a CMB-only correlation coefficient, by marginalizing over the foreground parameters.

\subsection{Cosmological results}\label{subsec:cosmo}
We use \texttt{cobaya} to sample the likelihoods and derive the posterior of the cosmological and foregrounds parameters. We run MCMC chains on Planck PR4 data using the $C_\ell$-likelihood and the $\mathcal{R}_\ell^{\rm TE}$-likelihood. We use wide flat priors for cosmological parameters, and a Gaussian prior on the optical depth of reionization $\tau = 0.054\pm 0.007$~\cite{Planck2018:cosmo}. The derivation of the statistical properties of the correlation coefficient (cf. appendix~\ref{app:statistics}) relies on a second order development, which is valid only on scales with a high $EE$ signal-to-noise ratio. We apply cuts in multipole to avoid the foreground-dominated scales (low-$\ell$) and noise dominated scales (high-$\ell$). The multipole ranges used in the two likelihoods for the different cross-frequency spectra are displayed in table~\ref{tab:multipoles}.\\

The posterior distributions for $H_0$, and the density parameters $\Omega_bh^2$ and $\Omega_ch^2$ are shown in Fig.~\ref{fig:cosmo}. We display the expected 1 $\sigma$ fluctuations between $C_\ell$-derived parameters and $\mathcal{R}_\ell^{TE}$-derived ones in gray. This was derived from the set of simulations introduced in section~\ref{sec:syst}. Cosmological parameters estimated from the correlation coefficient $\mathcal{R}_\ell^{TE}$ are consistent with the one estimated from power spectra. We do not detect the effect of any multiplicative systematic in the data.\\

\begin{table}
\begin{ruledtabular}
\begin{tabular}{lcc}
Parameter & $C_\ell^{\rm TT}$, $C_\ell^{\rm TE}$, $C_\ell^{\rm EE}$ & $\mathcal{R}_\ell^{\rm TE}$ \\
\hline
$H_0$ [km/s/Mpc] & $67.3 \pm 0.5$ & $67.5 \pm 1.3$\\
$\Omega_bh^2$ & $0.02233 \pm 0.00014$ & $0.02235 \pm 0.00037$\\
$\Omega_ch^2$ & $0.1194 \pm 0.0012$ & $0.1192 \pm 0.0028$\\
$\mathrm{ln}(10^{10}A_s)$ & $3.040 \pm 0.015$ & $3.098 \pm 0.152$
\end{tabular}
\end{ruledtabular}
\caption{\label{tab:results} Cosmological parameter constraints (mean values and $1\sigma$ errors) derived from the $C_\ell$-likelihood and the $\mathcal{R}_\ell^{\rm TE}$-likelihood.}
\end{table}

\begin{figure}[t]
    \centering
    \includegraphics{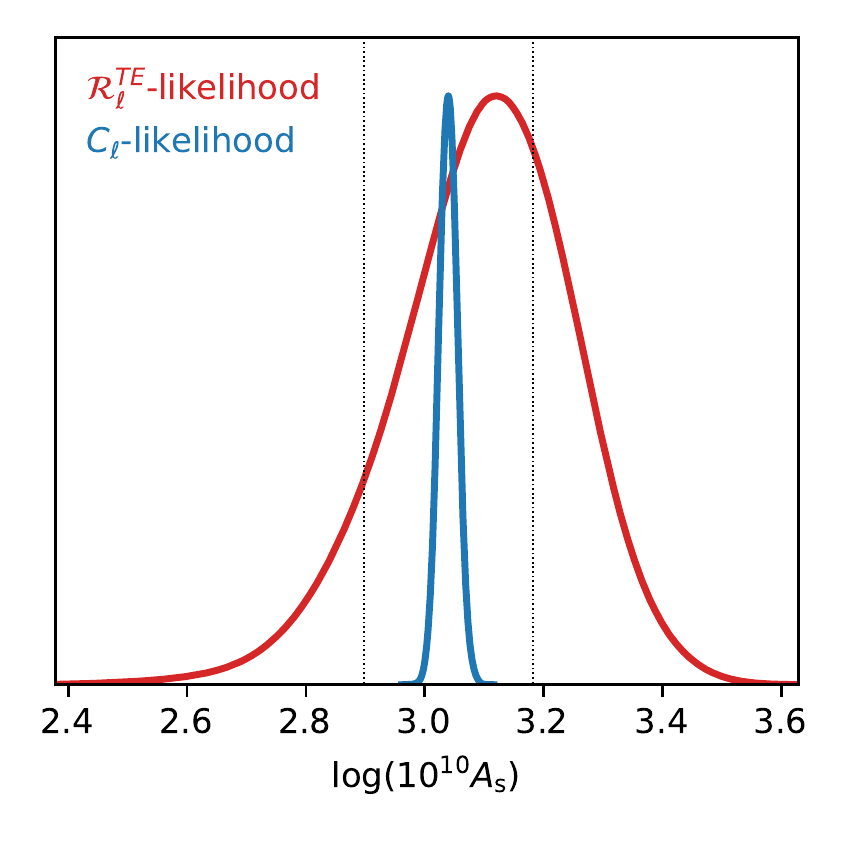}
    \caption{1D posterior distribution for $\mathrm{log}(10^{10}A_s)$ using the $C_\ell$-based likelihood (blue) or the $\mathcal{R}_\ell^{\rm TE}$-based likelihood (red). The correlation coefficient constraint is sensitive to $A_s$ only through the effect of lensing on the power spectra.}
    \label{fig:logA}
\end{figure}

Table~\ref{tab:results} presents mean values and 1$\sigma$ errors for the cosmological parameters discussed in this paper. As expected, using $\mathcal{R}_\ell^{\rm TE}$ worsens the constraints on cosmological parameters with respect to the $C_\ell$-based likelihood. The $\mathcal{R}_\ell^{\rm TE}$-derived errors on $H_0$ and $\Omega_bh^2$ are $2.6$ times wider than the $C_\ell$-derived ones and the error on $\Omega_ch^2$ is $2.3$ times larger. The correlation coefficient is poorly sensitive to the parameters describing the shape of the initial matter power spectrum, $A_s$ and $n_s$. Interestingly, while the amplitude parameter $A_s$ appears to cancel in the ratio of power spectra, it can be measured through the effect of lensing on the power spectra~\cite{Lewis2006}. The posterior distribution of $\mathrm{log}(10^{10}A_s)$ is displayed in Fig.~\ref{fig:logA}. The $\mathcal{R}_\ell^{\rm TE}$-derived error on $\mathrm{ln}(10^{10}A_s)$ is 10 times larger than the $C_\ell$-derived one. \\

We note that our value of the Hubble parameter $H_0 = 67.5 \pm 1.3 \;\mathrm{km/s/Mpc}$, derived from the correlation coefficient is consistent with other CMB based measurements of $H_0$. The Hubble parameter we obtain using $\mathcal{R}_\ell^{\rm TE}$ is still $3.1\sigma$ discrepant with the latest measurement from Cepheids-calibrated cosmic distance ladder~\cite{Riess2021}.

\subsection{CMB-only $\mathcal{R}_\ell^{TE}$}\label{subsec:cmbonly}
\begin{figure*}
\includegraphics[scale = 0.9]{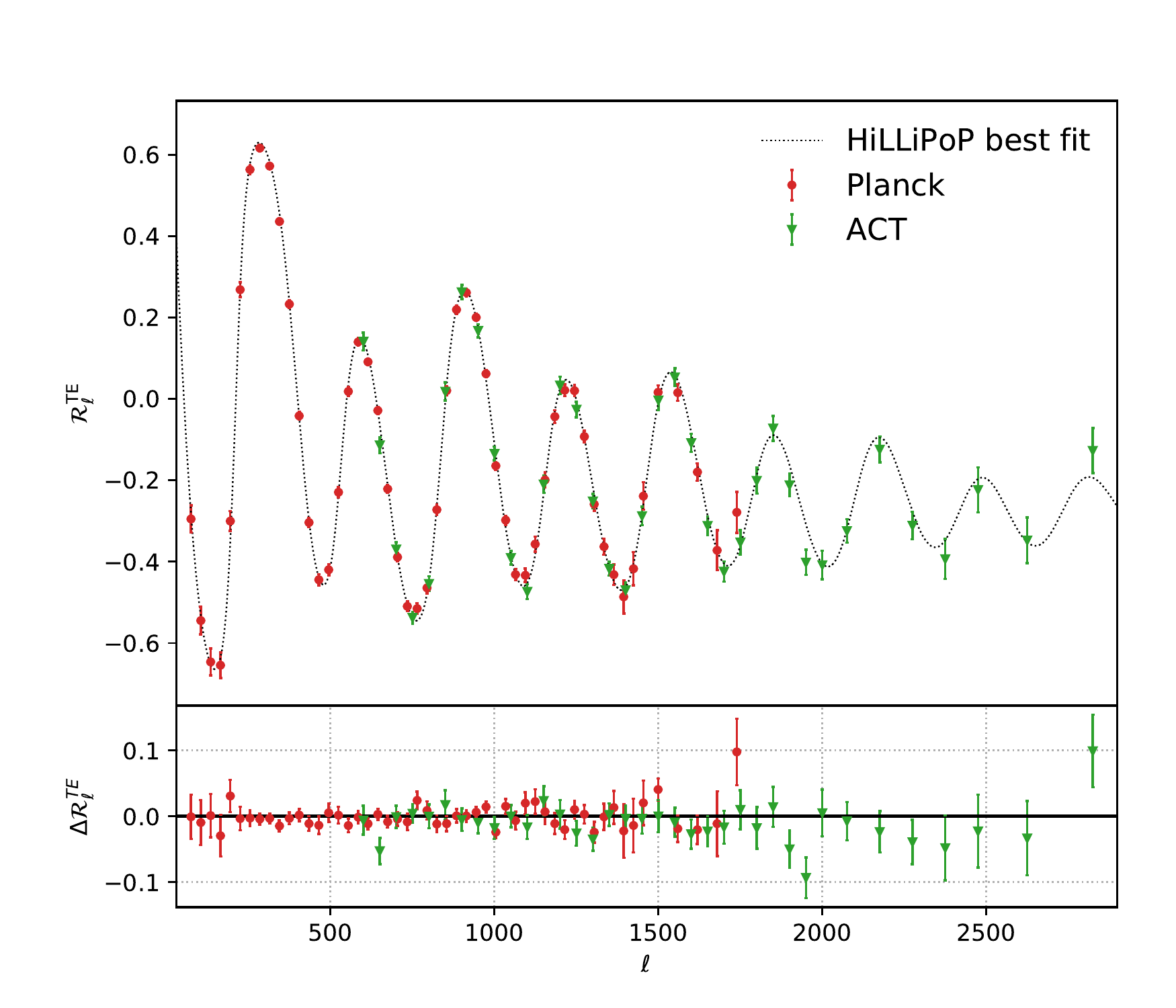}
\caption{\label{fig:cmb_only} (Top) : CMB-only $\mathcal{R}_\ell^{\rm{TE}}$ correlation coefficient for Planck PR4 data (red) and ACT data (green) after foreground marginalization. The best-fit correlation coefficient (gray dashed line) is computed with \texttt{CAMB} from the \texttt{HiLLiPoP} $C_\ell$-likelihood best fit cosmology --- (Bottom) : Residual plot with respect to the binned correlation coefficient best-fit.}
\end{figure*}

We produce CMB-only power spectra, marginalizing over foreground parameters. We follow the method used in~\cite{Dunkley2013}. The multi-frequency vector is modelled as
\begin{equation}\label{eq:cmb_model}
    C_\mathrm{b}^{\mathrm{model}} = \mathbf{A}\cdot C_\mathrm{b}^{\mathrm{CMB}} + C_\mathrm{b}^{\mathrm{fg}}(\theta_{\mathrm{fg}})
\end{equation}
with $\mathbf{A}$ a matrix that projects the CMB bandpowers to their corresponding elements in each cross-frequency spectrum, and $C_\mathrm{b}^{\mathrm{fg}}(\theta_{\mathrm{fg}})$ the foregrounds power spectra corresponding to a set of foreground parameters $\theta_\mathrm{fg}$.\\

Instead of sampling simultaneously CMB bandpowers and nuisance parameters, we split the joint probability distribution $p(C_\mathrm{b}^{\mathrm{CMB}}, \theta_\mathrm{fg}|\mathrm{data})$ into two conditional probability densities $p(C_\mathrm{b}^{\mathrm{CMB}}|\theta_\mathrm{fg},\mathrm{data})$ and $p(\theta_\mathrm{fg}|C_\mathrm{b}^{\mathrm{CMB}}, \mathrm{data})$. The main advantage here is that the CMB bandpowers sampling is simple because $C_\mathrm{b}^\mathrm{CMB}$ follows a multivariate normal distribution with mean $\hat{C}_\mathrm{b}$ and covariance $\mathbf{Q}$ given by

\begin{align}
    &\hat{C}_\mathrm{b} = [\mathbf{A}^\mathrm{T}\mathbf{\Sigma}^{-1}\mathbf{A}]^{-1}[\mathbf{A}^\mathrm{T}\mathbf{\Sigma}^{-1}(C_\mathrm{b}^\mathrm{data} - C_\mathrm{b}^\mathrm{fg}(\theta_\mathrm{fg}))]\label{eq:gibbs}\\
    &\mathbf{Q}^{-1} = \mathbf{A}^\mathrm{T}\mathbf{\Sigma}^{-1}\mathbf{A}\label{eq:covgibbs}
\end{align}
with $\mathbf{\Sigma}$ the $C_\ell$s multifrequency covariance matrix. We alternate CMB bandpowers sampling (using the known probability distribution described in Eqs.~[\ref{eq:gibbs},\ref{eq:covgibbs}]) and Metropolis-Hastings sampling for the nuisance parameters using a Gaussian likelihood. We then compute the correlation coefficient and its covariance matrix using the CMB-only power spectra and the $\mathbf{Q}$ covariance matrix.\\

In Fig.~\ref{fig:cmb_only}, we compare the CMB-only $\mathcal{R}_\ell^{\rm TE}$ with the best-fit model obtained using the $C_\ell$-based likelihood. The correlation coefficient is compatible with the model with $\chi^2/\mathrm{dof} = 52.2/52$ (PTE = 0.47).\\

Higher resolution experiments such as Atacama Cosmology Telescope (ACT) bring more information about the correlation coefficient at small scales. We display on Fig~\ref{fig:cmb_only} the ACT CMB-only $\mathcal{R}_\ell^{\rm TE}$, computed from the power spectra provided by the ACT collaboration~\cite{Choi2020} and publicly available.\footnote{\url{https://github.com/ACTCollaboration/}} Before computing the correlation coefficient, we remove points for which the $EE$ SNR is lower than $3$. We find that the ACT correlation coefficient is in good agreement with the Planck PR4 best-fit, with $\chi^2/\mathrm{dof} = 39.0/36$ (PTE = 0.33). 

\section{Conclusion}\label{sec:Conclusion}
Given the current context of tensions between early and late Universe measurements, it is important to assess the robustness of the cosmological parameter constraints. In this work, we have studied the impact of remaining multiplicative systematic effects that would have been neglected in the data treatment, resulting in a bias at the power spectra level. We have shown that this kind of bias can significantly shift the cosmological parameter constraints determined from the combination of $C_\ell^{\rm TT}$, $C_\ell^{\rm TE}$ and $C_\ell^{\rm EE}$.\\

To prevent the cosmological parameter measurements from being biased, we have proposed to use a likelihood based on the Pearson's correlation coefficient $\mathcal{R}_\ell^{\rm TE}$, an observable that is insensitive to multiplicative biases. We have obtained the first constraints on cosmology only from the correlation between $T$ and $E$ modes. We have shown that it gives cosmological parameters that are consistent with other CMB-based measurements. Using $\mathcal{R}_\ell^{\rm TE}$ increases the error on the cosmological parameters with respect to the errors derived from the combination of $TT$, $TE$, and $EE$ power spectra. However, we have measured the Hubble parameter $H_0 = 67.5 \pm 1.3$ km/s/Mpc with an associated error which is similar to the error obtained by ground-based CMB experiments such as ACTPol~\cite{ACT:A20}. The correlation coefficient provides a good consistency check on Planck data. We have not observed the effect of any multiplicative bias in Planck PR4 data and we have obtained a value for the Hubble parameter which is still discrepant with late Universe measurements.\\

Next generation of ground based CMB experiments such as Simons Observatory~\cite{Ade2019} or CMB-S4~\cite{Abazajian2016} will produce data with a high EE signal-to-noise ratio for a wide range of multipoles. This will allow the computation of $\mathcal{R}_\ell^{\rm TE}$ up to smaller scales, increasing its constraining power and relevance. Ground-based experiments are also typically more affected by transfer functions due to atmospheric and ground pick up filtering. We expect that observables such as $\mathcal{R}_\ell^{\rm TE}$ will provide a strong consistency check on these data.\\

\subsection*{Acknowledgements}
We thank Gilles Weymann for useful comments and discussion. The theoretical power spectra used in this paper were computed using \texttt{CAMB} Boltzmann solver~\cite{Lewis2000, Howlett2012}. We gratefully acknowledge the IN2P3 Computer Center (\url{http://cc.in2p3.fr}) for providing the computing resources and services needed to this work.

\onecolumngrid
\begin{appendix}

\section{Statistical properties of $\mathcal{R}_\ell^{\rm TE}$ estimator}\label{app:statistics}

In the high signal-to-noise regime, the correlation coefficient estimator can be developed as :
\begin{align}\label{eq:lin_dev}
    \hat{\mathcal{R}}_b^{\rm TE} &= \mathcal{R}_b^{\rm TE}\frac{1+ \frac{\Delta C_b^{\rm TE}}{C_b^{\rm TE}}}{\sqrt{\left(1+\frac{\Delta C_b^{\rm TT}}{C_b^{\rm TT}}\right)\left(1+\frac{\Delta C_b^{\rm EE}}{C_b^{\rm EE}}\right)}}\nonumber\\
    &= \mathcal{R}_b^{\rm TE}\left(1 + \frac{\Delta C_b^{\rm TE}}{C_b^{\rm TE}}\right)\cdot\left(1 - \frac{1}{2}\frac{\Delta C_b^{\rm TT}}{C_b^{\rm TT}} + \frac{3}{8}\left(\frac{\Delta C_b^{\rm TT}}{C_b^{\rm TT}}\right)^2\right)\cdot\left(1 - \frac{1}{2}\frac{\Delta C_b^{\rm EE}}{C_b^{\rm EE}} + \frac{3}{8}\left(\frac{\Delta C_b^{\rm EE}}{C_b^{\rm EE}}\right)^2\right)
\end{align}
where $\mathcal{R}_b^{\rm TE}$, $C_b^{\rm TT}$, $C_b^{\rm EE}$ and $C_b^{\rm TE}$ are the spectra we want to estimate.\\

Such an estimator is not an unbiased estimator (i.e. $\langle\hat{\mathcal{R}}_b^{\rm TE}\rangle \neq \mathcal{R}_b^{\rm TE}$), but is affected by a bias term such that : $\langle\hat{\mathcal{R}}_b^{\rm TE}\rangle = \mathcal{R}_b^{\rm TE}(1+\alpha_b)$. The bias term $\alpha_b$ can be easily computed taking the mean value of Eq.~(\ref{eq:lin_dev}) :
\begin{align}\label{eq:bias_cc}
    \alpha_b = \frac{3}{8}\left(\frac{\mathrm{cov}(C_b^{\rm TT}, C_b^{\rm TT})}{{C_b^{\rm TT}}^2} + \frac{\mathrm{cov}(C_b^{\rm EE}, C_b^{\rm EE})}{{C_b^{\rm EE}}^2}\right) - \frac{1}{2}\left(\frac{\mathrm{cov}(C_b^{\rm TT}, C_b^{\rm TE})}{C_b^{\rm TT}C_b^{\rm TE}} + \frac{\mathrm{cov}(C_b^{\rm EE}, C_b^{\rm TE})}{C_b^{\rm EE}C_b^{\rm TE}}\right) + \frac{1}{4}\frac{\mathrm{cov}(C_b^{\rm TT}, C_b^{\rm EE})}{C_b^{\rm TT}C_b^{\rm EE}}.
\end{align}

From Eq.~(\ref{eq:lin_dev}) we can compute the $\mathcal{R}_\ell^{TE}$ covariance matrices defined by
\begin{equation}\label{eq:r_covmat}
    \mathrm{cov}(\mathcal{R}_b^{\mathrm{TE}, \nu_1\times\nu_2},
                 \mathcal{R}_{b'}^{\mathrm{TE}, \nu_3\times\nu_4}) =
        \langle\Delta\mathcal{R}_b^{\mathrm{TE}, \nu_1\times\nu_2}
                \Delta\mathcal{R}_b^{\mathrm{TE}, \nu_3\times\nu_4}
        \rangle,
\end{equation}
with $\nu_1\times\nu_2$ and $\nu_3\times\nu_4$ two cross frequencies and $\Delta\mathcal{R}_b^{\rm TE} = \hat{\mathcal{R}}_b^{\rm TE} - \mathcal{R}_b^{\rm TE}$ the deviation to the mean value. $\Delta\mathcal{R}_b^{\rm TE}$ can be expressed analytically (at second order) using Eq.~(\ref{eq:lin_dev})

\section{\texttt{HiLLiPoP} power spectra model}\label{app:ps_model}
In section~\ref{sec:syst} we present the \texttt{HiLLiPoP} $C_\ell$-based likelihood. In section~\ref{sec:like}, we construct a $\mathcal{R}_\ell^{\rm TE}$-based likelihood. In this appendix, we explain how the model in Eq.~(\ref{eq:model_like}) is computed.\\

We work with three frequencies : 100, 143 and 217 GHz (indexed by $\nu_i$, $\nu_j$ for $(i, j)\in\{0,1,2\}^2$) and two split maps (indexed by $R$, $S$ with $(R,S)\in\{A,B\}^2$. We also define $X$ and $Y$ such as $(X,Y)\in\{T, E\}^2$. For two frequencies ($\nu_i$, $\nu_j$) with $j\ge i$ and two maps ($R$, $S$), we model the power spectra as

\begin{equation}\label{eq:ps_map}
    C_\ell^{XY, \mathrm{th}, \nu_i^R\times\nu_j^S} = A_{\rm Pl}c^{\nu_i^R}c^{\nu_j^S}\left[C_\ell^{XY, \mathrm{CMB}} + C_\ell^{XY, \mathrm{fg}, \nu_i\times\nu_j}\right],
\end{equation}
where $A_{\rm Pl}$ is a global amplitude calibration parameter and $c^{\nu_i^R}$, $c^{\nu_j^S}$ are calibration parameters at map level. This set of parameter is sampled in the likelihood as nuisance parameters.\\

We take the weighted average of cross-map power spectra to compute the cross-frequency power spectra, ignoring the auto power spectra.  

\begin{equation}\label{eq:ps_freq}
    C_\ell^{XY, \mathrm{th}, \nu_i\times\nu_j} = A_{\rm Pl}\left[\sum_{\substack{(R,S)\in\\\{A,B\}^2}} w_\ell^{XY, \nu_i^R\times\nu_j^S}c^{\nu_i^R}c^{\nu_j^S}\left(C_\ell^{XY, \mathrm{CMB}} + C_\ell^{XY, \mathrm{fg},\nu_i\times\nu_j}\right)\left(1 - \delta_{\nu_i\nu_j}(1 - \delta_{RA}\delta_{SB})\right)\right],
\end{equation}
where $w_\ell^{XY,\nu_i^R\times\nu_j^S}$ are the weights associated to the $\nu_i^R\times\nu_j^S$ $XY$ power spectrum and $\delta$ is the Kronecker delta. We can express Eq.~(\ref{eq:ps_freq}) as Eq.~(\ref{eq:model_like}) using the following definition

\begin{align}\label{eq:fg_freq}
    A^{XY, \nu_i\times\nu_j} = \sum_{\substack{(R,S)\in\\\{A,B\}^2}} w_\ell^{XY, \nu_i^R\times\nu_j^S}c^{\nu_i^R}c^{\nu_j^S}\left(1 - \delta_{\nu_i\nu_j}(1 - \delta_{RA}\delta_{SB})\right). 
\end{align}
We obtain the equation describing the model used to compute the power spectra in the likelihood
\begin{equation}
    C_\ell^{XY, \mathrm{th},\nu_i\times\nu_j} = A_\mathrm{Pl}A^{XY, \nu_i\times\nu_j}\left[C_\ell^{XY, \rm CMB} + C_\ell^{XY, \mathrm{fg}, \nu_i\times\nu_j}\right].
\end{equation}

\end{appendix}
\twocolumngrid
\bibliography{draft}

\end{document}